\documentclass[prd,aps,twocolumn]{revtex4}
\usepackage{natbib}
\usepackage{graphicx}
\usepackage{color}
\usepackage{hyperref}
\def\gr{$\gamma$-ray}

\begin{document}

\title{Radio-to-gamma-ray synchrotron and neutrino emission from proton-proton interactions in active galactic nuclei}

\author{Andrii Neronov$^{1,2}$ and Dmitri Semikoz$^{1,3,4}$ }
\affiliation{$^1$Université de Paris, CNRS, Astroparticule et Cosmologie,  F-75006 Paris, France\\
$^2$Astronomy Department, University of Geneva, Ch. d'Ecogia 16, 1290, Versoix, Switzerland
$^3$ Institute for Nuclear Research of the Russian Academy of Sciences, 60th October Anniversary Prospect 7a, Moscow 117312, Russia\\
$^4$National Research Nuclear University MEPHI (Moscow Engineering Physics Institute),
Kashirskoe highway 31, 115409 Moscow, Russia}

\begin{abstract}
We explore possible physical origin of correlation between radio wave and very-high-energy neutrino emission in active galactic nuclei (AGN), suggested by recently reported evidence for correlation between neutrino arrival directions and positions of brightest radio-loud AGN. We show that such correlation is expected if both synchrotron emitting electrons and neutrinos originate from decays of charged pions produced in  proton-proton interactions in parsec-scale relativistic jet propagating through circum-nuclear medium of the AGN.  
\end{abstract}
\maketitle

IceCube telescope has discovered astrophysical neutrino signal of uncertain origin in the very-high energy (VHE) range between TeV and 10 PeV \cite{icecube_science,icecube20}. Neutrinos with such energies can be generated in the Milky Way galaxy \cite{galactic,galactic1} and in extragalactic sources like radio-loud active galactic nuclei (AGN) \cite{biermann,neronov02,sibiryakov,ribordy,icecube_0506_2015}. IceCube data suggest that multiple spectral components might be present in the signal \cite{icecube20}. Low statistics of neutrino signal and moderate angular resolution of IceCube complicate identification of the source class responsible for the neutrino signal. 

Models of neutrino production in AGN conventionally assume that the neutrino flux is generated in decays of charged pions produced by interactions of  high-energy protons accelerated close to the supermassive black hole or in the relativistic jet ejected by the black hole. This process also inevitably produces electrons and \gr s with energies comparable to those of neutrinos. Hence it is natural to expect that neutrino-bright AGN should also be equally \gr\ bright \cite{biermann,neronov02,sibiryakov,ribordy,icecube_0506_2015}. A subtle point of this argument is that contrary to neutrinos which directly escape from the AGN source, \gr s and electrons do not directly escape. Instead, they can initiate development of electromagnetic cascade in the source \cite{biermann,mannheim}. The cascade transfers the power of electromagnetic emission toward lower energy range. As a result, the unambiguous relation between the neutrino and \gr\ power is lost. In addition, conventional models of \gr\ activity of AGN also consider signal from electrons accelerated in the jets. \gr\ emission of "leptonic" origin can be much stronger than that of "hadronic" origin. 
In fact, cross-correlation of the neutrino arrival directions with the positions of the \gr\ emitting AGN gives negative result and indicates that the \gr\ brightest AGN are not responsible for the astrophysical neutrino flux \cite{ptitsyna17,icecube_blazars}.

The "hadronic" models of AGN high-energy activity conventionally assume that protons mostly interact with low energy photons present in the AGN source \cite{biermann,mannheim,neronov02}. This process is characterised by high energy threshold for the pion production. Only protons with energies in excess of 1-10 PeV can produce neutrinos, \gr s and electrons in interactions with the optical-ultraviolet photon backgorund produced by the accretion flow. Electrons and \gr s generated by this process have energies in 10-100 TeV range. Electrons with such energies do not produce synchrotron emission in the radio band and there is no particular reason to expect strong correlation between the VHE neutrino and radio synchrotron flux.  

In what follows we consider an alternative scenario for neutrino production \cite{sibiryakov,ribordy}. In this scenario high-energy protons interact with low energy protons from the circum-nuclear regions around the AGN central engine. This process is characterised by the energy threshold in the GeV range and neutrinos, \gr s and electrons originating from this process can be injected with energies as low as 100~MeV. Electrons of such energies generate radio synchrotron emission. This might explain radio -- neutrino flux correlation suggested by analysis of Ref. \cite{plavin,plavin1}, where a correlation between the arrival directions of astrophysical neutrinos and radio-brightest AGN has been observed. 

We consider a high-energy proton beam 
ejected into jet with opening solid angle $\Omega$ along the direction of the AGN jet from acceleration region possibly close to the AGN central engine.  High-energy proton distribution is assumed to be a powerlaw  
\begin{equation}
    dN_p/dE\propto E^{-p}
\end{equation} with the slope $p$ close to $2-2.5$, expected from the shock acceleration process and consistent with the slope of the astrophysical neutrino signal \cite{icecube20}.

The beam propagates through  "circum-nuclear" medium of the AGN. The jet most probably escapes through a low density funnel in the direction aligned with the rotation axis of the black hole \cite{igumenshchev}. Even though the density profile of the funnel is uncertain, one can assume that it follows a powerlaw of the distance $r$    
\begin{equation}
    n(R)=n_0\left(\frac{r}{R_g}\right)^{-\gamma}
\end{equation}
where $R_g=G_NM\simeq 1.5\times 10^{13}\left[M/10^8M_\odot\right]$~cm is the gravitational radius of the central black hole of the mass $M$ and $G_N$ is the Newton's constant. The powerlaw index $\gamma$ can vary in wide range. It is $\gamma=1/2$ radiatively inefficient accretion flows \cite{riaf}, $\gamma=3/2$ for the spehrically symmetric Bondi accretion \cite{bondi}, $\gamma=2$ in the accretion disk winds \cite{winds}. 

The optical depth of the accreting material with respect to proton-proton interactions is determinet by the column density of material and by the the inelastic $pp$ interaction cross-section $\sigma_{pp}\sim 3\times 10^{-26}$~cm$^2$:
\begin{eqnarray}
&&\tau_{pp}=\sigma_{pp}\int_{R_g}^r n(r') dr'=\frac{\sigma_{pp}n_0 R_g}{1-\gamma}\left[\left(\frac{r}{R_g}\right)^{1-\gamma}-1\right]\nonumber\\
&&\simeq 1 \left[\frac{n_0}{10^{10}\mbox{ cm}^{-3}}\right]\left[\frac{M}{10^8M_\odot}\right]^{1/2}\left[\frac{r}{1\mbox{ pc}}\right]^{1/2}
\end{eqnarray}
(we adopt   $\gamma=1/2$ in this and subsequent  numerical estimates). The integral is dominated by the small $r$ limit if $\gamma<1$, e.g. in the case of wind-like density profile with $\gamma=2$. 

\begin{figure}
    \includegraphics[width=\linewidth]{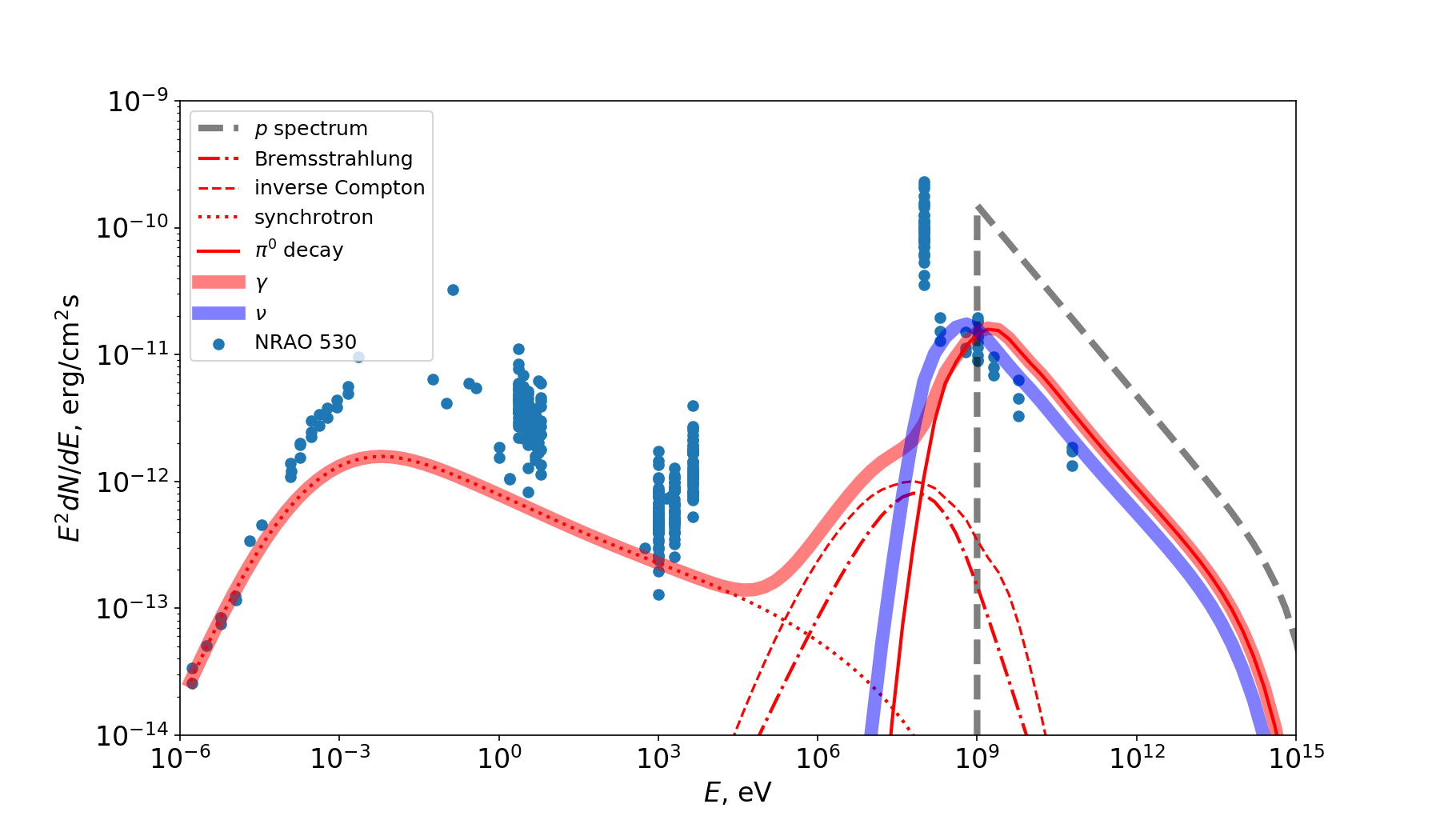}         
    \includegraphics[width=\linewidth]{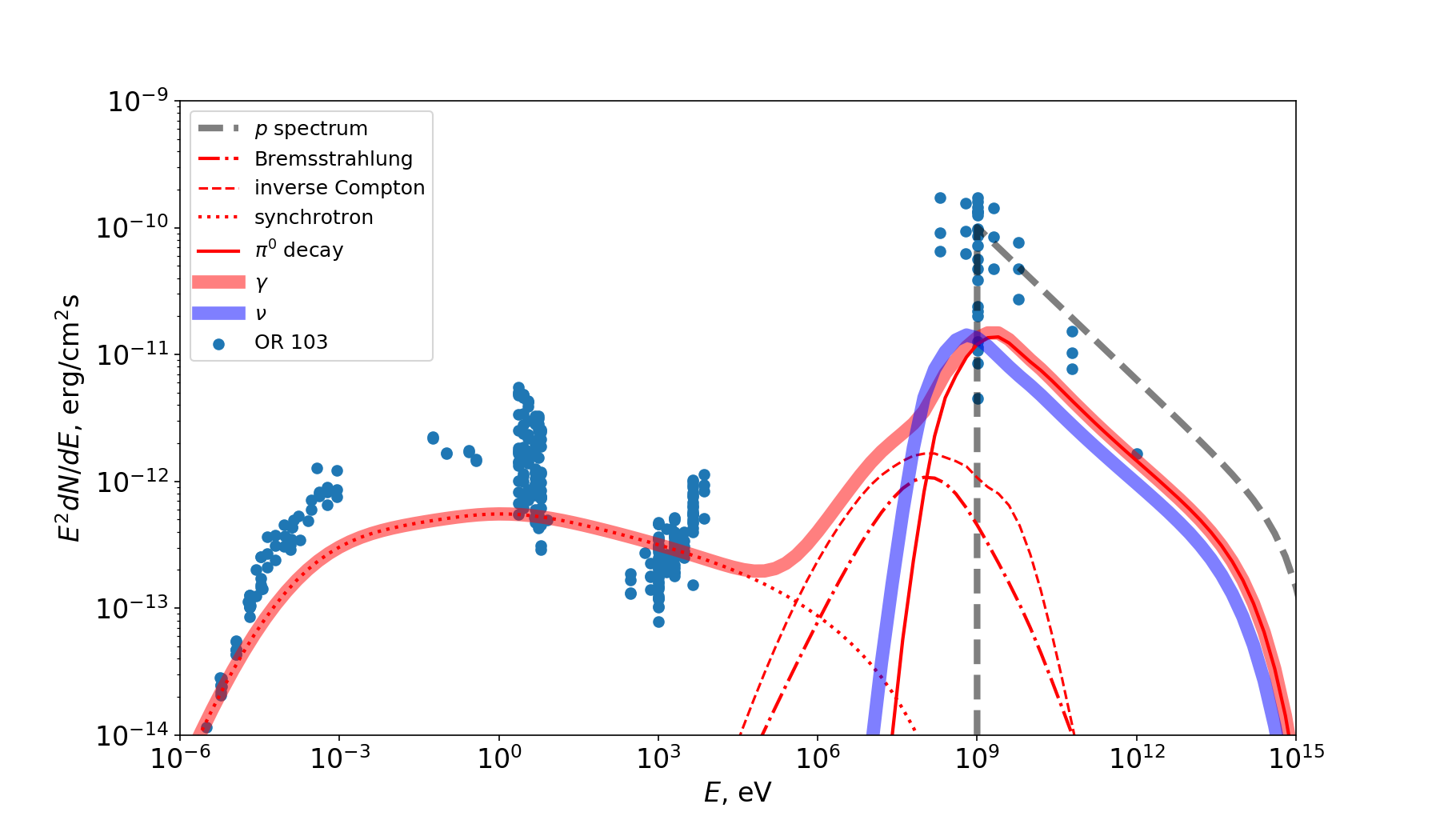}
    \includegraphics[width=\linewidth]{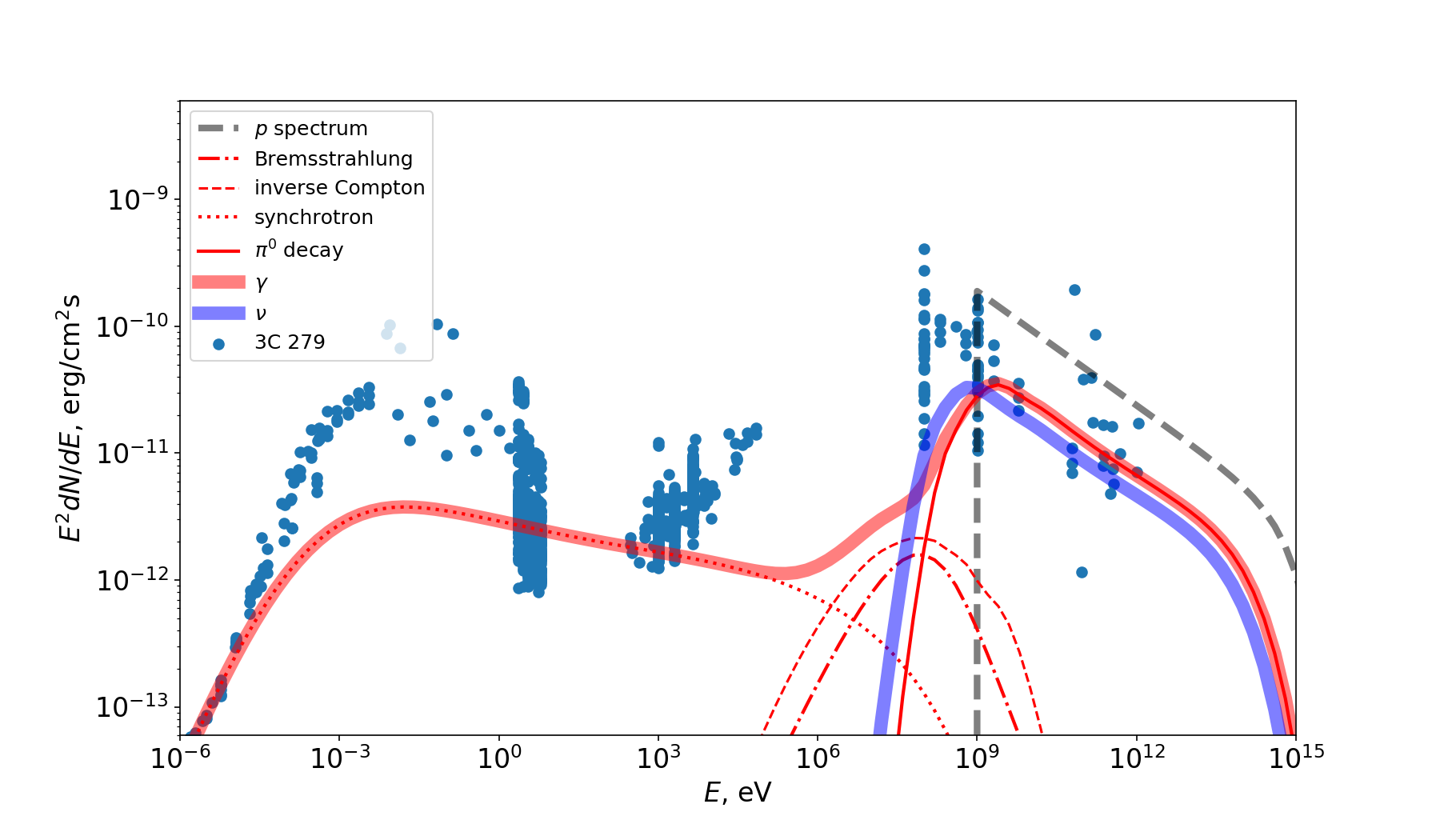}   \includegraphics[width=\linewidth]{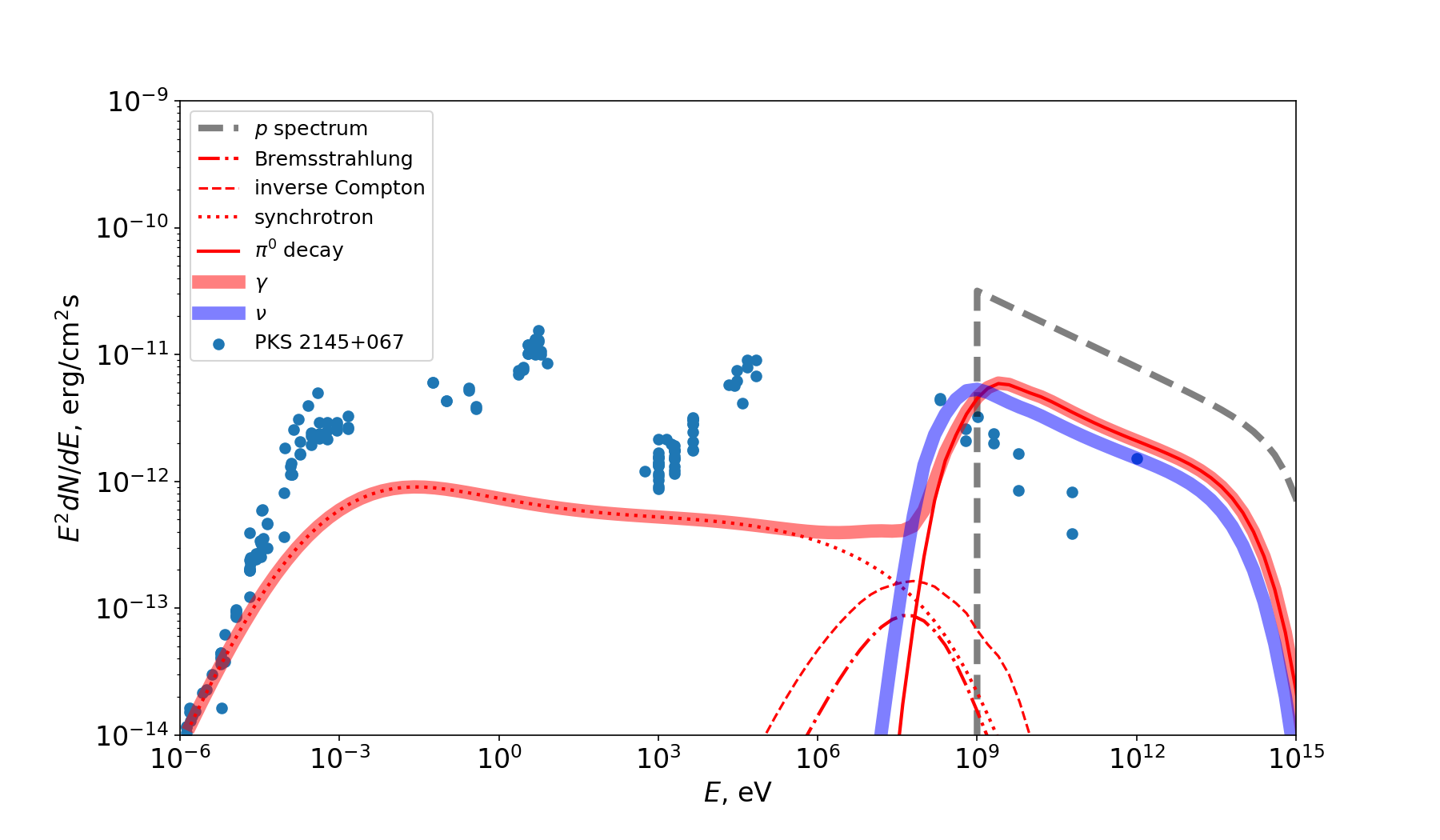}
\caption{Spectra of neutrino (blue) and electromagnetic (red) emission from $pp$ interactions. Data point show  spectral energy distributions of electromagnetic radiation from  PKS 2145+067, NRAO 530, 3C 379 and OR 103 (from top to bottom) retreived using the SED builder {\tt https://tools.ssdc.asi.it/SED/}. Data for NRAO 530 are from  \cite{1999ApJS..123...79H,2015ApJS..218...23A,2012ApJS..199...31N,2010ApJS..188..405A,2016A&A...588A.103B,1999A&A...349..389V,0067-0049-210-1-8,2013A&A...551A.142D,2010AJ....140.1868W,2015arXiv150702058P,2014A&A...571A..28P,2011A&A...536A...7P,1994ApJS...91..111W,1990PKS...C......0W,1998AJ....115.1693C,1970ApJS...20....1D,2007ApJS..171...61H}. Data for OR 103 are from \cite{0004-637X-779-1-27,2012ApJS..199...31N,2010ApJS..188..405A,0067-0049-210-1-8,2013A&A...551A.142D,2012A&A...541A.160G,2011MNRAS.411.2770B,2010AJ....140.1868W,2009ApJS..180..283W,2015arXiv150702058P,2014A&A...571A..28P,1992ApJS...79..331W,1996ApJS..103..427G,2011A&A...536A...7P,1990PKS...C......0W,1998AJ....115.1693C,1981A&AS...45..367K,2007MNRAS.376..371J,1970ApJS...20....1D,2007ApJS..171...61H,2003MNRAS.341....1M}. For PKS 2145+067 the data are from \cite{0004-637X-779-1-27,2013ApJS..207...19B,2012ApJS..199...31N,2010ApJS..188..405A,2010A&A...524A..64C,2010A&A...510A..48C,2016A&A...588A.103B,1999A&A...349..389V,0067-0049-210-1-8,2013A&A...551A.142D,2012A&A...541A.160G,2011MNRAS.411.2770B,2010AJ....140.1868W,2009ApJS..180..283W,2015arXiv150702058P,2014A&A...571A..28P,1992ApJS...79..331W,1996ApJS..103..427G,2011A&A...536A...7P,1994ApJS...91..111W,1990PKS...C......0W,1998AJ....115.1693C,1981A&AS...45..367K,2007MNRAS.376..371J,1970ApJS...20....1D,2007ApJS..171...61H,2003MNRAS.341....1M}. For 3C 279 the data are from \cite{0004-637X-779-1-27,1999ApJS..123...79H,2009A&A...506.1563P,2013ApJS..207...19B,2010ApJS..186....1B,2015ApJS..218...23A,2012ApJS..199...31N,2010ApJS..188..405A,2012ApJ...749...21A,2010A&A...524A..64C,2010A&A...510A..48C,2015arXiv150407051R,2008A&A...480..611S,2016A&A...588A.103B,1999A&A...349..389V,1992ApJS...80..257E,0067-0049-210-1-8,2013A&A...551A.142D,2012A&A...541A.160G,2010AJ....140.1868W,2009ApJS..180..283W,2015arXiv150702058P,2014A&A...571A..28P,1992ApJS...79..331W,2011A&A...536A...7P,2010MNRAS.402.2403M,1990IRASF.C......0M,1994ApJS...91..111W,1990PKS...C......0W,1998AJ....115.1693C,1981A&AS...45..367K,1997ApJ...475..479W,1970ApJS...20....1D,2007ApJS..171...61H,2002babs.conf...63G}.}
    \label{fig:plavin}
\end{figure}

Proton-proton interactions result in production and decays of neutral and charged pions. The spectra of \gr s, electrons and neutrinos  produced in pion  decays can be calculated using a range of publicly available codes providing parameterisations of the differential production cross-sections  \cite{kelner,kamae,kafeixu,aafrag}. Fig. \ref{fig:plavin}  shows the spectra of \gr\ and neutrino emission from pion production and decays in $pp$ interactions calculated using the parameterisation of \cite{kamae} at the energies below 10 GeV and those of \cite{aafrag} at the energies above 10 GeV. We use such a combination because these two different parametersations are specifically tuned for lower and higher energy ranges. The parametersation of \cite{kamae} available through a public library {\it cparamlib} explicitly ignores primary protons with energies above 400 TeV. To the contrary, parameterisation of \cite{aafrag} are valid for protons with energies above  $4$~GeV. The slopes of the proton spectra are $p=2.5$ for NRAO 530 case, $p=2.4$ for OR103, $p=2.3$ for 3C 279 and $p=2.2$ for PKS 2145+067.

The $\pi^0$ decay \gr\ spectrum and $\pi^\pm$ decay neutrino spectra are cut-off powerlaw types with similar normalisation and slopes close to the slope of the parent high-energy proton spectrum.

The neutrino spectra can hardly me measured because the flux from individual sources is only marginally detectable by IceCube, so that non-negligible signal statistics has to be accumulated from a source population, rather than on source-by-source basis \cite{neronov_jetp}. This approach has actually been adopted in the analysis of neutrino -- radio flux correlation \cite{plavin,plavin1}.

The \gr\ spectra shown in Fig. \ref{fig:plavin} are modified  by the effect of attenuation on Extragalactic Background Light \cite{franceschini08} (not shown in the figure). This is a relatively well controlled modification of the \gr\ spectrum. Another modification of the \gr\ spectrum by the pair production is expected because of the development of electromagnetic cascade in the radiation field of the source. A quantitative prediction of this effect on the spectrum is hardly possible because this requires knowledge of the spectral properties of distance-dependent soft photon backgrounds in the source over a wide range of distances, from the vicinity of the black hole to parsec-scale distances. This effect is moderately important if most of $pp$ interactions occur at pc-scale distances (this is the case for shallow density profiles like $\gamma=1/2$). If the column density for $pp$ interactions is accumulated close to the black hole (e.g. for wind-like density profile with $\gamma=2$), the intrinsic pair production is more important and has strong influence on the source \gr\ spectrum. 

Apart from \gr s and neutrinos, pion production and decay results also in production of electrons which are not directly detectable, but contribute to the source luminosity once they loose energy via one of the radiative cooling mechanisms: synchrotron radiation, inverse Compton scattering of Bremsstrahlung emission. 

The strength and spectral characteristics  of the inverse Compton  emission are determined by the same distance-dependent soft photon background in the source and it is equally difficult to establish. Inverse Compton scattering is an essential part of the electromagnetic cascade process supported by the $\gamma\gamma$ pair production converting \gr s into electrons and positrons and by the inverse Compton scattering regenerating \gr s. 

A quantitative estimate of the influence of the inverse Compton scattering on electrons can be obtained in the Thomson regime, below the energy threshold of electromagnetic cascade. In this regime the inverse Compton emission power is determined by the energy density of the soft photon backgorund providing seed photons for Compton scattering. Assuming that a fraction $\kappa_{rad}$ of radiation from the central source of luminosity $L\sim 10^{45}$~erg/s is scattered in the medium,  one can find that the density of photon background contributing to the inverse Compton scattering at a distance $r$: $U_{rad}\sim \kappa_{rad} L/(4\pi r^2)$
\begin{eqnarray}
    &&U_{rad}\simeq  2\times 10^{8}\kappa_{rad}\left[\frac{L}{10^{45}\mbox{ erg/s}}\right] \left[\frac{r}{1\mbox{ pc}}\right]^{-2}\frac{\mbox{eV}}{\mbox{cm}^3}
\end{eqnarray}
Electrons of energy $E_e$ loose energy on inverse Compton scattering on the time scale
\begin{eqnarray}
    &&t_{ics}\simeq \frac{3m_e^2}{4U_{rad} E_e}\simeq \frac{2\times  10^{9}}{\kappa_{rad}}\left[\frac{L}{10^{45}\mbox{ erg/s}}\right]^{-1}\\
    &&\left[\frac{r}{1\mbox{ pc}}\right]^2\left[\frac{E_e}{10^{8}\mbox{ eV}}\right]^{-1}\mbox{ s}\nonumber
\end{eqnarray}
If the typical energy of photons of the low-energy photon backgorund is $\epsilon\sim 10$~eV (e.g. ultraviolet radiation from the accretion disk scattered in the Broad-Line Region), the Thomson regime of Compton scattering applies to electrons producing \gr s with energies below 
\begin{equation}
\label{eq:KN}
    E_{\gamma,T}=\frac{m_e^2}{\epsilon}\simeq 25\left[\frac{\epsilon}{10\mbox{ eV}}\right]^{-1}\mbox{ GeV}
\end{equation}
For such electrons relative importance of the synchrotron and inverse Compton losses is determined by the ratio of the energy densities of magnetic field and radiation. Fig. \ref{fig:plavin} shows the result of calculation of the synchrotron and inverse Compton emission from pion decay electrons, computed using the code of Ref. \cite{zdz} for the central source with $L=10^{45}$~erg/s, found for a black body source with $T=1.4\times 10^5$~K and radius $R_*=3\times 10^{13}$~cm. It is interesting to notice that the code of Ref. \cite{zdz} has been originally developed to describe interactions of pulsar wind with the radiation and wind of a massive star. As a matter of fact, modification of parameters (the luminosity of the central source, the central density and slope of the matter radial profile can be directly used to describe propagation of proton beam through the AGN environment.

Apart from Compton scattering, electrons also suffer from synchrotron energy loss. Its importance is  determined by the strength of magnetic field. 
An estimate of the magnetic field can be obtained from the equipartition argument. Assuming that the magnetic field energy density is a fraction $\kappa_m$ of the kinetic energy density of matter at the distance $R$ moving with velocities close to the free-fall velocity scale $v\sim \sqrt{R_g/r}$, one finds 
\begin{eqnarray}
\label{eq:B}
    &&B= \left(\frac{8\pi\kappa_m  n_0 R_g^{\gamma+1}}{r^{\gamma+1}}\right)^{1/2}\simeq 0.1 \left[\frac{\kappa_m}{0.01}\right]^{1/2} \\ &&\left[\frac{M}{10^8M_\odot}\right]^{3/4}\left[\frac{r}{1\mbox{ pc}}\right]^{-3/4}\left[\frac{n_0}{10^{10}\mbox{ cm}^{-3}}\right]^{1/2}\mbox{ G}\nonumber
\end{eqnarray}
where we have again used $\gamma=1/2$ for the numerical estimate. 

The rate of the synchrotron energy loss is determined by the energy density of magnetic field
\begin{eqnarray}
&&U_B=\frac{B^2}{8\pi}\simeq 3\times 10^{8}\left[\frac{\kappa_m}{0.01}\right]\\ &&\left[\frac{M}{10^8M_\odot}\right]^{3/2}\left[\frac{r}{1\mbox{ pc}}\right]^{-3/2}\left[\frac{n_0}{10^{10}\mbox{ cm}^{-3}}\right]\mbox{ eV/cm}^3\nonumber
\end{eqnarray}
The relative importance of the synchrotron and inverse Compton losses in Thomson regime is determined by the ratio of the energy densities of magnetic and radiation fields, $U_B/U_{rad}$. 

Fig. \ref{fig:plavin} shows the spectrum of synchrotron emission from the pion decay electrons calculated for the magnetic field radial profile is  $B=0.05[r/1\mbox{pc}]^{-3/4}$~G (3C279, NRAO 530 and PKS 2145+067 panels) or $B=0.025[r/1\mbox{pc}]^{-3/4}$~G (OR 103 panel) which corresponds to $\kappa_m\sim 0.005$ for the central density $n_0=10^{10}$~cm$^{-3}$ and radial profile with $\gamma=1/2$. 

Electrons injected by the pion decays are allowed to cool due to the synchrotron and inverse Compton loss (see \cite{zdz}) for details of the calculation). The injection spectrum of electrons produced in pion decays closely follows the spectra of photons and neutrinos. It also has the slope similar to the slope of the parent proton spectrum and a high-energy cut-off close to the cut-offs of the neutrino and \gr\ spectra. Similarly to the \gr\ and neutrino spectrum, the electron spectrum has a low-energy cut-off around the threshold of the pion production, somewhat below 100~MeV. 

Synchrotron photons generated by electrons of the energy $E_e$ have energies
\begin{eqnarray}
    &&\epsilon_s=\frac{eBE_e^2}{m_e^3}\simeq \ 5\times 10^{-5}\left[\frac{\kappa_m}{0.01}\right]^{1/2} \left[\frac{M}{10^8M_\odot}\right]^{3/4}\\ &&\left[\frac{r}{1\mbox{ pc}}\right]^{-3/4}\left[\frac{n_0}{10^{10}\mbox{ cm}^{-3}}\right]^{1/2}\left[\frac{E_e}{10^{8}\mbox{ eV}}\right]^2\mbox{ eV}\nonumber
\end{eqnarray}
If the electron spectrum extends as a powerlaw between $10^8$~eV and $10^{15}$~eV, over at least seven decades in energy, the synchrotron spectrum is expected to extend over at least fourteen decades in energy up to 
$\epsilon_s\simeq 5\left[B/0.1\mbox{ G}\right]\left[E_e/10^{15}\mbox{ eV}\right]^2\mbox{ GeV}$
from radio band up to the \gr\ band. 

The synchrotron cooling time is 
\begin{eqnarray}
    &&t_s=\frac{3m_e^4}{4e^4B^2E_e}\simeq 2\times 10^8\left[\frac{\kappa_m}{0.01}\right]^{-1}\left[\frac{M}{10^8M_\odot}\right]^{-3/2} \\ &&\left[\frac{r}{1\mbox{ pc}}\right]^{3/2}\left[\frac{n_0}{10^{10}\mbox{ cm}^{-3}}\right]^{-1}
    \left[\frac{E_e}{10^{8}\mbox{ eV}}\right]^{-1}\mbox{ s}\nonumber
\end{eqnarray}
Comparing the synchrotron cooling time to the dynamical time scale $t_{esc}=r/c\simeq  10^8\left[r/1\mbox{ pc}\right]\mbox{ s}$ we find that 100 MeV electrons originating from pion decays efficiently cool within the distance range where $t_s<t_{esc}$, i.e. within the distance
\begin{eqnarray}
    &&r_s=1\left[\frac{\kappa_m}{0.01}\right]^{1/2}\left[\frac{M}{10^8M_\odot}\right]^{3/4} \\ &&\left[\frac{r}{1\mbox{ pc}}\right]^{-3/4}\left[\frac{n_0}{10^{10}\mbox{ cm}^{-3}}\right]^{1/2}
    \left[\frac{E_e}{10^{8}\mbox{ eV}}\right]^{1/2}\mbox{ pc}\nonumber
\end{eqnarray}
in the case of $\gamma=1/2$ density profile with the central density $n_0\sim 10^{10}$~cm$^{-3}$.

The spectrum of electrons injected from pion decays within the distance $r\lesssim r_s$  is softened by the effect of synchrotron cooling. It is a powerlaw  
\begin{equation}
    \frac{dn_s}{d\epsilon_s}\propto \epsilon_s^{-\gamma_s}, \ \ \ \gamma_s=\frac{p_e+1}{2}=\frac{p}{2}+1
\end{equation}
If the spectrum of protons in the beam has the slope $p=2$, the spectrum of synchrotron emission also has the slope $\gamma_s\simeq 2$ over a broad energy range from radio to \gr s.

Modelling shown in Fig. \ref{fig:plavin} demonstrates that, as expected, the synchrotron emission component extends over wide energy range. The radio synchrotron emission is generated by electrons with energies close to the pion production threshold $E_e\sim 100$~MeV. It is efficiently produced together with neutrino flux at the distances $R\gtrsim 10^{17}$~cm where the magnetic field is low enough and the synchrotron emission energy decreases into the radio range. This is the distance scale at which radio jets from AGN can be resolved by the Very-Long Baseline Interferometry (VLBI) technique. 
Both the radio synchrotron and neutrino fluxes are directly proportional to the power of the proton beam. This can readily explains the correlation of the radio and neutrino emission power reported in Ref. \cite{plavin}.

Apart from the synchrotron and inverse Compton emission, emission, electrons propagating through the medium inevitably loose energy onto Bremsstrahlung. The Bremsstahlung energy loss time of Bremsstrahlung
can be comparable to the synchrotron loss time for the 100~MeV energy electrons close to the pion production threshold. The energy of the Bremsstrahlung photons is comparable to the energy of electrons. The Bremsstrahlung spectrum nearly repeats the parent electron spectrum (since the cooling time does not depend on energy). This explains the presence of non-negligible Bremsstrahlung emission component in the spectra shown in Fig. \ref{fig:plavin} in the 100 MeV range.

Model calculations of Fig. \ref{fig:plavin} are done for shallow density profile $\gamma=1/2$. In this case bulk of the neutrino and electromagnetic flux from $pp$ interactions is generated at pc-scale distances. At such distances $\gamma\gamma$ pair production does not modify the electromagnetic spectrum. To the contrary, $\gamma\gamma$ pair production initiates the development of electromagnetic cascade and modification of the source spectrum in the case $\gamma>1$. In this case bulk of the neutrino flux is produced close to the black hole, rather than at pc-scale distances. Electromagnetic cascade removes power from the highest energy range and can transfer it to the energy range just below the pair production threshold. This happens if electrons and positrons generated by the pair production process do not loose energy preferentially onto synchrotron emission. As discussed above, the relative importance of the inverse Compton scattering compared to the synchrotron is regulated by the energy densities of magnetic and radiation fields. Uncertainties of details of electromagnetic cascade development  strongly reduce the robustness of calculations of expected power of the source in the \gr\ band. This complicates the analysis of possible correlation between the neutrino and \gr\ emissivities. 

A significant uncertainty of both radio -- neutrino and  \gr\ -- neutrino flux correlations is that of the relative importance of the "leptonic" component of \gr\ and radio flux. Conventional models of high-energy activity of AGN assume that both the radio synchrotron and \gr\ inverse Compton emission are produced by electrons directly accelerated in the AGN jets, rather than produced in pion decays. Radiation from the pion decay electrons and from the pion decay \gr\ induced cascade form the "hadronic" flux component. In general,  both components co-exist and their relative contributions have to be determined based on the modelling of the spectral energy distribution of the source. Fig. \ref{fig:plavin} shows that in any case, only part of electromagnetic flux from the AGN contributing to the radio-neutrino correlation considered in \cite{plavin} can be attributed to the radiation directly originating from pion decays.  The rest has to be attributed either to the pion decay induced electromagnetic cascade or to the "leptonic" emission from  directly accelerated primary electrons (rather than from secondary electrons from pion decays).  
\section*{Acknowledgement}

This work is supported by the Ministry of science and higher education of Russian Federation under the 
contract 075-15-2020-778 in the framework of the Large scientific projects program within the national project "Science".

\bibliography{references}
\end{document}